\documentclass[onecolumn,showpacs,showkeys,amsmath,amssymb,superscriptaddress,floatfix,nofootinbib]{revtex4}

\usepackage{graphicx}
\usepackage{bm} 
\usepackage{subfigure}

\makeatletter
\newcommand\erfc{\mathop{\operator@font erfc}\nolimits}
\def\slashchar#1{\setbox0=\hbox{$#1$}
   \dimen0=\wd0 \setbox1=\hbox{/} \dimen1=\wd1
   \ifdim\dimen0>\dimen1 \rlap{\hbox to \dimen0{\hfil/\hfil}} #1
   \else  \rlap{\hbox to \dimen1{\hfil$#1$\hfil}} / \fi}

\begin{document}
 
\title{
Projection method and new formulation of leading-order anisotropic hydrodynamics 
}

\author{Leonardo Tinti} 
\email{tinti@fi.infn.it}
\affiliation{Institute of Physics, Jan Kochanowski University, PL-25406~Kielce, Poland}

\author{Wojciech Florkowski} 
\email{Wojciech.Florkowski@ifj.edu.pl}
\affiliation{Institute of Physics, Jan Kochanowski University, PL-25406~Kielce, Poland} 
\affiliation{The H. Niewodnicza\'nski Institute of Nuclear Physics, Polish Academy of Sciences, PL-31342 Krak\'ow, Poland}

\date{December 23, 2013}

\begin{abstract}
The introduced earlier projection method for boost-invariant and cylindrically symmetric systems is used to introduce a new formulation of anisotropic hydrodynamics that allows for three substantially different values of pressure acting locally in three different  directions. Our considerations are based on the Boltzmann kinetic equation with the collision term treated in the relaxation time approximation and the momentum anisotropy is included explicitly in the leading term of the distribution function. A novel feature of our work is the complete analysis of the second moment of the Boltzmann equation, in addition to the zeroth and first moments that have been analyzed in earlier studies. We define the final equations of anisotropic hydrodynamics in the leading order as a subset of the analyzed moment equations (and their linear combinations) which agree with the Israel-Stewart theory in the case of small pressure anisotropies.
\end{abstract}



\maketitle 

%%%%%%%%%%%%%%%%%%%%%%%%%%%%%%%%%%%%%%%%%%%%%%%%%%%%%%%%%%%%%%%%%%%%%%%%%%%%%%%%%%%%%%%%%%%%%%%%%%%%
\section{Introduction}
\label{sect:intro}
%%%%%%%%%%%%%%%%%%%%%%%%%%%%%%%%%%%%%%%%%%%%%%%%%%%%%%%%%%%%%%%%%%%%%%%%%%%%%%%%%%%%%%%%%%%%%%%%%%%%%

Successful applications of relativistic viscous hydrodynamics in the description of heavy-ion collisions at RHIC (Relativistic Heavy-Ion Collider) and the LHC (Large Hadron Collider) triggered large interest in the development of the hydrodynamic framework
\cite{Israel:1976tn,Israel:1979wp,
Muronga:2001zk,Muronga:2003ta,
Baier:2006um,Baier:2007ix,
Romatschke:2007mq,Dusling:2007gi,Luzum:2008cw,
Song:2008hj,El:2009vj,PeraltaRamos:2010je,
Denicol:2010tr,Denicol:2010xn,
Schenke:2010rr,Schenke:2011tv,
Bozek:2009dw,Bozek:2011wa,
Niemi:2011ix,Niemi:2012ry,
Bozek:2012qs,Denicol:2012cn,Jaiswal:2013npa}. An example of the new approach to relativistic dissipative hydrodynamics is {\it anisotropic hydrodynamics} \cite{Florkowski:2010cf,Martinez:2010sc,
Ryblewski:2010bs,Martinez:2010sd,
Ryblewski:2011aq,Martinez:2012tu,
Ryblewski:2012rr,Ryblewski:2013jsa,
Florkowski:2012ax,Florkowski:2012as} --- the framework where effects connected with the expected high pressure anisotropy of the produced matter are included in the leading order of the hydrodynamic expansion. Very recently, also the second order anisotropic hydrodynamics has been formulated by Bazow, Heinz, and Strickland~\cite{Bazow:2013ifa}. The~new approach introduced in \cite{Bazow:2013ifa} allows for description of arbitrary transverse expansion of matter in the way which becomes consistent with more traditional approaches to dissipative hydrodynamics in the small anisotropy limit. This formalism  uses, however, the Romatschke-Strickland form \cite{Romatschke:2003ms} of the distribution function  in the leading order, which implies that the two components of pressure in the transverse plane may be different only if the second-order corrections are taken into account.

In this work we present a new methodology for including three substantially different pressure components already in the leading order of hydrodynamic expansion. Our approach is based on the projection method introduced in Ref.~\cite{Florkowski:2011jg}, which has turned out to be a convenient tool to replace complicated tensor equations of relativistic hydrodynamics by a small set of scalar equations. We take into account the radial expansion of the produced matter (in addition to the longitudinal Bjorken flow) but our considerations are confined to the case with cylindrical symmetry. We generalize the Romatschke-Strickland form to the case where all three pressure components may be different. Compared to earlier works on anisotropic hydrodynamics in the leading order, where the zeroth and first moments of the Boltzmann equation have been studied, an important novel feature of our present work is the analysis of the second moment of the Boltzmann equation. We argue that a successful agreement with the Israel-Stewart theory in the limit of small anisotropies may be achieved if we take into account two equations constructed from the second moment of the Boltzmann equation rather than taking one equation from the zeroth moment and another equation from the second moment. 

In our opinion, the use of the second moment sheds new light on the framework of anisotropic hydrodynamics. We expect, that the formalism developed in this paper may be a better starting point for the second-order anisotropic hydrodynamics developed according to the guidelines presented in Ref.~\cite{Bazow:2013ifa}. In addition,  the presented approach may be generalized in the natural way to the 2+1 case where the cylindrical symmetry is relaxed.

The paper is organized as follows: In the next Section we introduce the four-vectors $U$, $X$, $Y$, and $Z$ used to decompose different tensors used in our formalism, in particular, to decompose the expansion and shear tensors. In Sec.~\ref{sect:BE} we discuss the Boltzmann equation in the relaxation time approximation and introduce the anisotropic distribution function characterized by three anisotropy parameters. The zeroth moment of the Boltzmann equation is discussed shortly in Sec.~\ref{sect:0mom}. In Sec.~\ref{sect:1mom} we characterize the energy-momentum conservation law, the Landau matching condition, and the close-to-equilibrium limit of the energy-momentum tensor. The formulas for the energy-density and pressure of anisotropic systems are presented in Sec.~\ref{sect:enedenaniso}. Sec.~\ref{sect:2mom} contains the analysis of the second moment of the Boltzmann equation. The most important part of the paper, Sec.~\ref{sect:set}, describes the construction of two equations (out of the complete set of second moment equations) which are finally accepted as the two new equations of anisotropic hydrodynamics in the leading order. The entropy production and its positivity is discussed in Sec.~\ref{sect:ent}. We summarize and conclude in Sec.~\ref{sect:con}. Two appendices containing explicit forms of different expressions and integrals close the paper. Throughout the paper we use natural units where $c=\hbar=k_B=1$ and the metric tensor with the signature $(+,-,-,-)$.

%%%%%%%%%%%%%%%%%%%%%%%%%%%%%%%%%%%%%%%%%%%%%%%%%%%%%%%%%%%%%%%%%%%%%%%%%%%%%%%%%%%%%%%%%%%%%%%%%%%%%
%%%%%%%%%%%%%%%%%%%%%%%%%%%%%%%%%%%%%%%%%%%%%%%%%%%%%%%%%%%%%%%%%%%%%%%%%%%%%%%%%%%%%%%%%%%%%%%%%%%%%
\section{Projection method for boost-invariant and cylindrically symmetric hydrodynamic systems}
\label{sect:projection}
%%%%%%%%%%%%%%%%%%%%%%%%%%%%%%%%%%%%%%%%%%%%%%%%%%%%%%%%%%%%%%%%%%%%%%%%%%%%%%%%%%%%%%%%%%%%%%%%%%%%%
%%%%%%%%%%%%%%%%%%%%%%%%%%%%%%%%%%%%%%%%%%%%%%%%%%%%%%%%%%%%%%%%%%%%%%%%%%%%%%%%%%%%%%%%%%%%%%%%%%%%%

%%%%%%%%%%%%%%%%%%%%%%%%%%%%%%%%%%%%%%%%%%%%%%%%%%%%%%%%%%%%%%%%%%%%%%%%%%%%%%%%%%%%%%%%%%%%%%%%%%%%%
%%%%%%%%%%%%%%%%%%%%%%%%%%%%%%%%%%%%%%%%%%%%%%%%%%%%%%%%%%%%%%%%%%%%%%%%%%%%%%%%%%%%%%%%%%%%%%%%%%%%%
\subsection{Boost-invariant and cylindrically symmetric flow}
\label{sect:flow}
%%%%%%%%%%%%%%%%%%%%%%%%%%%%%%%%%%%%%%%%%%%%%%%%%%%%%%%%%%%%%%%%%%%%%%%%%%%%%%%%%%%%%%%%%%%%%%%%%%%%%
%%%%%%%%%%%%%%%%%%%%%%%%%%%%%%%%%%%%%%%%%%%%%%%%%%%%%%%%%%%%%%%%%%%%%%%%%%%%%%%%%%%%%%%%%%%%%%%%%%%%%

The space-time coordinates and the four-vector describing the hydrodynamic flow are denoted in the standard way as 
$x^\mu = \left( t, x, y, z \right)$ and 
\begin{equation}
U^\mu = \gamma (1, v_x, v_y, v_z), \quad \gamma = (1-v^2)^{-1/2}.
\label{Umu0}
\end{equation}
For boost-invariant and cylindrically symmetric systems, the scalar quantities may depend only on the (longitudinal) proper time and the radial distance
\begin{equation}
\tau = \sqrt{t^2 - z^2}, \quad r = \sqrt{x^2 + y^2}.
\label{taur}
\end{equation} 
In addition, for the boost-invariant hydrodynamic flow (\ref{Umu0}) we may use the following parametrization
\begin{eqnarray}
U^0 = \cosh \theta_\perp \cosh \eta_\parallel, \quad
U^1 = \sinh \theta_\perp \cos  \phi,          
\quad 
U^2 = \sinh \theta_\perp \sin  \phi,           \quad
U^3 = \cosh \theta_\perp \sinh \eta_\parallel,
\label{Umu}
\end{eqnarray}
where $\theta_\perp=\theta_\perp(\tau,r)$ is the transverse fluid rapidity defined by the formula
\begin{equation}
v_\perp = \sqrt{v_x^2+v_y^2} = \frac{\tanh \theta_\perp}{\cosh\eta_\parallel}.
\label{thetaperp}
\end{equation} 
Here $\eta_\parallel$ is the space-time rapidity
and $\phi$ is the azimuthal angle
\begin{eqnarray}
\eta_\parallel = \frac{1}{2} \ln \frac{t+z}{t-z}, 
\quad \phi = \arctan \frac{y}{x}.
\label{etaparphi} 
\end{eqnarray}

In addition to $U^\mu$ we define three other four-vectors. The first one, $Z^\mu$, defines the longitudinal direction that plays a special role due to the initial geometry of the collision, 
\begin{eqnarray}
Z^0 = \sinh \eta_\parallel, \quad
Z^1 = 0,          \quad
Z^2 = 0,           \quad
Z^3 = \cosh \eta_\parallel.
\label{Zmu}
\end{eqnarray}
The second four-vector, $X^\mu$, defines a transverse direction to the beam,
\begin{eqnarray}
X^0 = \sinh \theta_\perp \cosh \eta_\parallel, \quad
X^1 = \cosh \theta_\perp \cos  \phi,           \quad
X^2 = \cosh \theta_\perp \sin  \phi,           \quad
X^3 = \sinh \theta_\perp \sinh \eta_\parallel,
\label{Xmu}
\end{eqnarray}
while the third four-vector, $Y^\mu$, defines the second transverse direction,
\begin{eqnarray}
Y^0 = 0, \quad
Y^1 = -\sin  \phi,           \quad
Y^2 = \cos  \phi,           \quad
Y^3 = 0.
\label{Ymu}
\end{eqnarray}

The four-vector $U^\mu$ is time-like, while the four-vectors $Z^\mu, X^\mu, Y^\mu$ are space-like. In addition, they are all orthogonal to each other, 
\begin{eqnarray}
U^2 &=& 1, \quad Z^2 = X^2 = Y^2 = -1, \nonumber \\
U \cdot Z &=& 0, \quad U \cdot X = 0, \quad U \cdot Y = 0, \nonumber \\
Z \cdot X &=& 0, \quad Z \cdot Y = 0, \quad X \cdot Y = 0.
\label{norm}
\end{eqnarray}
All these properties are most easily seen in the {\it local rest frame} of the fluid element (LRF), where we have \mbox{$\theta_\perp = \eta_\parallel = \phi = 0$} and
\begin{eqnarray}
U = (1,0,0,0),           \quad
Z = (0,0,0,1),           \quad
X = (0,1,0,0),           \quad
Y = (0,0,1,0).
\label{LRF}
\end{eqnarray}

In the standard formalism of dissipative hydrodynamics one uses the operator $ \Delta^{\mu \nu} = g^{\mu \nu} - U^\mu U^\nu$, that projects on the three-dimensional space orthogonal to $U^\mu$. It can be shown that 
\begin{equation}
\Delta^{\mu \nu} = g^{\mu \nu} - U^\mu U^\nu = -X^\mu X^\nu - Y^\mu Y^\nu - Z^\mu Z^\nu.
\label{Delta}
\end{equation}
Using Eqs. (\ref{norm}) we find that $Z^\mu, X^\mu$ and $Y^\mu$ are the eigenvectors of $\Delta^{\mu \nu}$,
\begin{equation}
\Delta^{\mu}_{\,\, \nu} \,X^\nu = X^\mu, \quad  \Delta^{\mu}_{\,\, \nu} \,Y^\nu = Y^\mu, \quad 
\Delta^{\mu}_{\,\, \nu} \,Z^\nu = Z^\mu.
\label{eigen}
\end{equation}
In this work, following the method of Ref.~\cite{Florkowski:2011jg}, we use the tensor products of the four-vectors $U, X, Y$, and $Z$ as the basis to decompose all other tensors appearing in the formalism of standard dissipative hydrodynamics and anisotropic hydrodynamics. This allows us to replace complicated tensor equations by a set of scalar equations and to identify the key degrees of freedom in anisotropic hydrodynamics. Various formulas and identities satisfied by the four-vectors $U, X, Y$, and $Z$, and also by their derivatives are listed in Sec.~\ref{sect:explicitr}. We shall refer frequently to those expressions in this paper. 

%%%%%%%%%%%%%%%%%%%%%%%%%%%%%%%%%%%%%%%%%%%%%%%%%%%%%%%%%%%%%%%%%%%%%%%%%%%%%%%%%%%%%%%%%%%%%%%%%%%%%
%%%%%%%%%%%%%%%%%%%%%%%%%%%%%%%%%%%%%%%%%%%%%%%%%%%%%%%%%%%%%%%%%%%%%%%%%%%%%%%%%%%%%%%%%%%%%%%%%%%%%
\subsection{Expansion and shear tensors}
\label{sect:expandshear}
%%%%%%%%%%%%%%%%%%%%%%%%%%%%%%%%%%%%%%%%%%%%%%%%%%%%%%%%%%%%%%%%%%%%%%%%%%%%%%%%%%%%%%%%%%%%%%%%%%%%%
%%%%%%%%%%%%%%%%%%%%%%%%%%%%%%%%%%%%%%%%%%%%%%%%%%%%%%%%%%%%%%%%%%%%%%%%%%%%%%%%%%%%%%%%%%%%%%%%%%%%%

For the sake of convenience, we present now explicit forms of the expansion and shear tensors expressing them in terms of $X$, $Y$ and $Z$. In the general case, the expansion tensor is defined by the formula~\cite{Muronga:2003ta}
\begin{equation}
\theta_{\mu \nu} = \Delta^\alpha_\mu \Delta^\beta_\nu \partial_{(\beta} U_{\alpha)},
\label{theta-munu}
\end{equation}
where the brackets denote the symmetric part of $\partial_{\beta} U_{\alpha}$. Using Eqs. (\ref{Umu}) in the definition of the expansion tensor (\ref{theta-munu}) and also using Eqs. (\ref{Zmu})--(\ref{Ymu}),  we find that the following decomposition holds for boost-invariant and cylindrically symmetric systems \cite{Florkowski:2011jg}~\footnote{We stress that the subscripts $X$, $Y$, and $Z$ do not denote the Cartesian coordinates but refer typically to the coefficients in the s such as Eq.~(\ref{theta-dec}).}
\begin{equation}
\theta^{\mu \nu} = \theta_X X^\mu X^\nu + \theta_Y Y^\mu Y^\nu +  \theta_Z Z^\mu Z^\nu, 
\label{theta-dec}
\end{equation}
where
\begin{equation}
\theta_X = - \frac{\partial \theta_\perp}{\partial r} \cosh \theta_\perp 
- \frac{\partial \theta_\perp}{\partial \tau} \sinh \theta_\perp, \quad
\theta_Y = - \frac{\sinh \theta_\perp}{r}, \quad
\theta_Z = - \frac{\cosh \theta_\perp}{\tau}.
\label{thetas}
\end{equation}
The contraction of the tensors $\Delta^{\mu \nu}$ and $\theta^{\mu \nu}$ gives the volume expansion parameter $\theta = \Delta^{\mu \nu} \theta_{\mu \nu}$. Equations~(\ref{Delta})--(\ref{theta-munu}) yield
\begin{eqnarray}
\theta = -\theta_X - \theta_Y - \theta_Z.
\label{volexpp}
\end{eqnarray} 
It is interesting to check that the volume expansion parameter $\theta$ may be expressed also by the formula $\theta = \partial_\mu U^\mu$. 

In addition to the expansion tensor $\theta^{\mu\nu}$ we shall use the shear tensor $\sigma_{\mu \nu}$. The latter is defined by the formula
\begin{equation}
\sigma_{\mu \nu} = \theta_{\mu \nu} - \frac{1}{3} \Delta_{\mu \nu} \theta.
\label{sigma1}
\end{equation}
With the help of the decompositions (\ref{Delta}) and (\ref{theta-dec}) we may write
\begin{equation}
\sigma^{\mu \nu} = \sigma_X X^\mu X^\nu + \sigma_Y Y^\mu Y^\nu +  \sigma_Z Z^\mu Z^\nu,
\label{sigma-dec}
\end{equation}
where
\begin{eqnarray}
\sigma_X &=& \frac{\theta}{3}+\theta_X = \frac{\cosh \theta_\perp}{3 \tau} + 
\frac{\sinh\theta_\perp}{3r} 
-\frac{2}{3} \frac{\partial\theta_\perp}{\partial \tau} \sinh\theta_\perp 
-\frac{2}{3} \frac{\partial\theta_\perp}{\partial r} \cosh\theta_\perp ,  \label{sigmaX}
\end{eqnarray}
\begin{eqnarray}
\sigma_Y &=& \frac{\theta}{3}+\theta_Y = \frac{\cosh \theta_\perp}{3 \tau} -
\frac{2 \sinh\theta_\perp}{3r} 
+\frac{1}{3} \frac{\partial\theta_\perp}{\partial \tau} \sinh\theta_\perp 
+\frac{1}{3} \frac{\partial\theta_\perp}{\partial r} \cosh\theta_\perp , \label{sigmaY}
\end{eqnarray}
and
\begin{eqnarray}
\sigma_Z &=& \frac{\theta}{3}+\theta_Z  
= -\frac{2\cosh \theta_\perp}{3 \tau} +
\frac{\sinh\theta_\perp}{3r} 
+\frac{1}{3} \frac{\partial\theta_\perp}{\partial \tau} \sinh\theta_\perp 
+\frac{1}{3} \frac{\partial\theta_\perp}{\partial r} \cosh\theta_\perp . \label{sigmaZ}
\end{eqnarray}
In agreement with general requirements we find that
\begin{eqnarray}
\sigma_X+\sigma_Y+\sigma_Z=0.
\label{sumsigma}
\end{eqnarray} 
In the case where the radial flow is absent \mbox{$\sigma_X = \sigma_Y = 1/(3 \tau)$} and \mbox{$\sigma_Z = -2/(3 \tau)$}, which agrees with earlier findings \cite{Muronga:2003ta}. For brevity of notation, expressions such as Eqs.~(\ref{theta-dec}) or (\ref{sigma-dec}) will be written shortly as the sums
\begin{eqnarray}
\theta^{\mu\nu} = \sum_I \theta_I I^\mu I^\nu,
\qquad
\sigma^{\mu\nu} = \sum_I \sigma_I I^\mu I^\nu,\end{eqnarray}
where $I$ takes the values $X$, $Y$, and $Z$. Similarly, Eqs.~(\ref{volexpp}) and (\ref{sumsigma}) may be written as
\begin{eqnarray}
\theta = -\sum_I \theta_I, \qquad
\sum_I \sigma_I = 0.
\end{eqnarray}

%%%%%%%%%%%%%%%%%%%%%%%%%%%%%%%%%%%%%%%%%%%%%%%%%%%%%%%%%%%%%%%%%%%%%%%%%%%%%%%%%%%%%%%%%%%%%%%%%%%%%
%%%%%%%%%%%%%%%%%%%%%%%%%%%%%%%%%%%%%%%%%%%%%%%%%%%%%%%%%%%%%%%%%%%%%%%%%%%%%%%%%%%%%%%%%%%%%%%%%%%%%
\section{Boltzmann equation and anisotropic distribution functions}
\label{sect:BE}
%%%%%%%%%%%%%%%%%%%%%%%%%%%%%%%%%%%%%%%%%%%%%%%%%%%%%%%%%%%%%%%%%%%%%%%%%%%%%%%%%%%%%%%%%%%%%%%%%%%%%
%%%%%%%%%%%%%%%%%%%%%%%%%%%%%%%%%%%%%%%%%%%%%%%%%%%%%%%%%%%%%%%%%%%%%%%%%%%%%%%%%%%%%%%%%%%%%%%%%%%%%

The basis for our considerations is the Boltzmann equation treated in the relaxation time approximation
\cite{Bhatnagar:1954zz,Baym:1984np,Baym:1985tna,
Heiselberg:1995sh,Wong:1996va}
\begin{equation}
 p\cdot\partial f =  \frac{p\cdot U}{\tau_{\rm eq}}
 \left(f_{\rm eq} - f \right).
 \label{RTA}
\end{equation}
In Eq.~(\ref{RTA}) $f$ is the phase-space distribution function, $f_{\rm eq}$ is the equilibrium distribution function, and $\tau_{\rm eq}$ is the relaxation time. In different frameworks of anisotropic hydrodynamics which have been studied so far, one assumes that the distribution function $f$ is very well approximated by the  Romatschke-Strickland form \cite{Romatschke:2003ms}. The use of this form is, however, not satisfactory in the cases where the transverse expansion is included in addition to the longitudinal Bjorken flow. If the effects connected with shear viscosity are taken into account, the presence of the transverse flow induces differences between the two components of pressure in the transverse plane. The Romatschke-Strickland form allows for the difference between the longitudinal and transverse pressures but the two transverse pressures must be identical.

An essential new feature of the present work is the generalization of the Romatschke-Strickland form to the expression which allows for three different components of pressure
\begin{eqnarray}
 f(x,p) = k \exp\left(-\frac{1}{\Lambda}\sqrt{ \left( 1 + \zeta_X \right) \left( p\cdot X \right)^2 + \left( 1 + \zeta_Y \right) \left( p\cdot Y \right)^2 + \left( 1 + \zeta_Z \right) \left( p\cdot Z \right)^2 }\right).
 \label{fzeta}
\end{eqnarray}
Here $k$ is an overall normalization constant, $\Lambda$ is the typical momentum scale, and $\zeta_I$'s $(I=X,Y,Z)$ are three anisotropy parameters. In the special case where $\zeta_X=\zeta_Y=0$, Eq.~(\ref{fzeta}) is reduced to the Romatschke-Strickland form \cite{Romatschke:2003ms}. In more general cases, the function (\ref{fzeta}) depends on the three different ratios  $(1+\zeta_I)/\Lambda^2$. Exactly this feature allows us to introduce three different components of pressure in the local rest frame.

Introducing the new variables, namely, 
\begin{eqnarray}
\lambda = \frac{ \Lambda }{ \sqrt{ 1 + 
\frac{ 1 }{ 3 } (\zeta_X + \zeta_Y + \zeta_Z)} }, \quad \xi_I =  \frac{1 +\zeta_I}{ 1 + \frac{1}{ 3 } ( \zeta_X + \zeta_Y + \zeta_Z )} - 1 
 \quad (I=X,Y,Z),
\end{eqnarray}
the distribution function (\ref{fzeta}) may be rewritten in the equivalent form as
\begin{eqnarray}
 f(x,p) &=& k \exp\left(-\frac{1}{\lambda}\sqrt{ \left( 1 + \xi_X \right) \left( p\cdot X \right)^2 + \left( 1 + \xi_Y \right) \left( p\cdot Y \right)^2 + \left( 1 + \xi_Z \right) \left( p\cdot Z \right)^2 }\,\,\right) \nonumber \\
 &=& k \exp\left(-\frac{1}{\lambda}\sqrt{ 
 \left( p\cdot U\right)^2 +
\xi_X \left( p\cdot X \right)^2 + 
\xi_Y \left( p\cdot Y \right)^2 + 
\xi_Z \left( p\cdot Z \right)^2 }\,\,\right),
\label{fxi}
\end{eqnarray}
where the new anisotropy parameters $\xi_I$ satisfy the condition 
\begin{eqnarray}
\sum_I \xi_I = \xi_X + \xi_Y + \xi_Z = 0.
\label{sumofxis}
\end{eqnarray}
To replace the first line in Eq.~(\ref{fxi}) by the second line we used Eq.~(\ref{Delta}) and the mass-shell condition $p^2=m^2=0$.  The physical constraints $\Lambda > 0$ and $1+\zeta_I > 0$ imply that $\lambda > 0$ and  $1+\xi_I > 0$. Hence, the initial parametrization (\ref{fzeta}) is completely equivalent to the new one. Below we shall use the expression (\ref{fxi}) and treat the scale $\lambda$ together with the two anisotropy parameters $\xi_X$ and $\xi_Y$ as three independent variables~\footnote{We note that the covariant form of the distribution function depends also on the transverse fluid rapidity $\theta_\perp$ through the vectors $U^\mu$ and $X^\mu$. Hence, we have in fact four independent scalar functions in (\ref{fxi}).}. Equation (\ref{sumofxis}) defines the applicability range of our parameterization 
\begin{eqnarray}
-1 < \xi_X, \quad -1 < \xi_Y, \quad \xi_X+\xi_Y < 1.
\label{range}
\end{eqnarray}
The equilibrium function in (\ref{RTA}) has the form
\begin{eqnarray}
f_{\rm eq}(x,p) = k \exp\left( - \frac{p \cdot U}{T}
\right).
\label{feq}
\end{eqnarray}
One can show that the distribution function (\ref{fxi}) is reduced to the form (\ref{feq}) with $\lambda=T$, if the anisotropy parameters $\xi_I$ are all set equal to zero.

%%%%%%%%%%%%%%%%%%%%%%%%%%%%%%%%%%%%%%%%%%%%%%%%%%%%%%%%%%%%%%%%%%%%%%%%%%%%%%%%%%%%%%%%%%%%%%%%%%%%%
%%%%%%%%%%%%%%%%%%%%%%%%%%%%%%%%%%%%%%%%%%%%%%%%%%%%%%%%%%%%%%%%%%%%%%%%%%%%%%%%%%%%%%%%%%%%%%%%%%%%%
\section{Zeroth moment and particle number density}
\label{sect:0mom}
%%%%%%%%%%%%%%%%%%%%%%%%%%%%%%%%%%%%%%%%%%%%%%%%%%%%%%%%%%%%%%%%%%%%%%%%%%%%%%%%%%%%%%%%%%%%%%%%%%%%%
%%%%%%%%%%%%%%%%%%%%%%%%%%%%%%%%%%%%%%%%%%%%%%%%%%%%%%%%%%%%%%%%%%%%%%%%%%%%%%%%%%%%%%%%%%%%%%%%%%%%%

In this Section we present the zeroth moment of the Boltzmann equation. The zeroth and the first moments of the Boltzmann equation were used in Refs.~\cite{Martinez:2010sc,Martinez:2010sd,
Martinez:2012tu} to derive equations of anisotropic hydrodynamics in the direct relation to kinetic theory. This approach is suitable for the analysis of one-dimensional boost-invariant flow, since the first two moments yield three equations for three unknown functions (in this work these functions have been introduced as $\Lambda$, $T$, and $\zeta_Z$). If the transverse flow is included, one has to take into consideration one or more equations from the second moment of the Boltzmann equation. Below, we shall argue that in the boost-invariant and cylindrically symmetric case (with non-zero radial flow) it is preferable to consider two equations from the second moment rather than one equation from the zeroth moment together with an extra equation obtained from the second moment. Consequently, the formulas introduced in this Section will serve only as the reference point.

Having in mind the comments stated above, we introduce the zeroth moment of the kinetic equation (\ref{RTA}) 
\begin{eqnarray}
\int\!\! dP \; p\cdot\partial f =  \frac{1}{\tau_{\rm eq}} \int\!\! dP \,  p\cdot U
 \left(  f_{\rm eq} -f \right).
\label{zm1}
\end{eqnarray}
Here $dP=d^3{\bf p}/p$ is the Lorentz invariant integration measure (for massless particles considered in this work \mbox{$p = \sqrt{p_x^2+p_y^2+p_z^2}$}). Using the standard definition of the particle number current we find
\begin{eqnarray}
N^\mu = \int dP\, p^\mu f = n \,U^\mu, \quad 
N^\mu_{\rm eq} = \int dP\, p^\mu f_{\rm eq}
= n_{\rm eq}\, U^\mu,
\end{eqnarray} 
and
\begin{eqnarray}
D n + n \theta = \frac{1}{\tau_{\rm eq}} \left(
n_{\rm eq} - n \right),
\label{zm2}
\end{eqnarray}
where $\theta$ is the expansion parameter defined in (\ref{volexpp}). We note that there are no terms proportional to the four-vectors $X^\mu$, $Y^\mu$ or $Z^\mu$ in the expansion of the current $N^\mu$. This is due to the quadratic dependence of the distribution function (\ref{fxi}) on these four-vectors. Dividing (\ref{zm2}) by $n$ we may further rewrite the zeroth moment equation as
\begin{eqnarray}
D \ln n + 
\theta = \frac{1}{\tau_{\rm eq}} 
\left( \frac{n_{\rm eq}}{n}-1\right).
\label{zm3}
\end{eqnarray}

The particle number density $n$ calculated for the anisotropic distribution function (\ref{fxi}) equals
\begin{eqnarray}
n(\lambda,\xi) = 
\frac{8\pi k \lambda^3}{\sqrt{1+\xi_X} \sqrt{1+\xi_Y}\sqrt{1+\xi_Z}}.
\label{n}
\end{eqnarray}
On the left-hand side of (\ref{n}) we use the short-hand notation, $\xi$, to denote three anisotropy parameters $\xi_X$, $\xi_Y$, and $\xi_Z=-\xi_X-\xi_Y$. In equilibrium, the expression for the particle number density simplifies to
\begin{eqnarray}
n_{\rm eq}(T) = 8\pi k T^3.
\label{neq}
\end{eqnarray}

%%%%%%%%%%%%%%%%%%%%%%%%%%%%%%%%%%%%%%%%%%%%%%%%%%%%%%%%%%%%%%%%%%%%%%%%%%%%%%%%%%%%%%%%%%%%%%%%%%%%%
%%%%%%%%%%%%%%%%%%%%%%%%%%%%%%%%%%%%%%%%%%%%%%%%%%%%%%%%%%%%%%%%%%%%%%%%%%%%%%%%%%%%%%%%%%%%%%%%%%%%%
\section{First moment of kinetic equation}
\label{sect:1mom}
%%%%%%%%%%%%%%%%%%%%%%%%%%%%%%%%%%%%%%%%%%%%%%%%%%%%%%%%%%%%%%%%%%%%%%%%%%%%%%%%%%%%%%%%%%%%%%%%%%%%%
%%%%%%%%%%%%%%%%%%%%%%%%%%%%%%%%%%%%%%%%%%%%%%%%%%%%%%%%%%%%%%%%%%%%%%%%%%%%%%%%%%%%%%%%%%%%%%%%%%%%%

%%%%%%%%%%%%%%%%%%%%%%%%%%%%%%%%%%%%%%%%%%%%%%%%%%%%%%%%%%%%%%%%%%%%%%%%%%%%%%%%%%%%%%%%%%%%%%%%%%%%%
%%%%%%%%%%%%%%%%%%%%%%%%%%%%%%%%%%%%%%%%%%%%%%%%%%%%%%%%%%%%%%%%%%%%%%%%%%%%%%%%%%%%%%%%%%%%%%%%%%%%%
\subsection{Energy-momentum conservation}
\label{sect:enmomcon}
%%%%%%%%%%%%%%%%%%%%%%%%%%%%%%%%%%%%%%%%%%%%%%%%%%%%%%%%%%%%%%%%%%%%%%%%%%%%%%%%%%%%%%%%%%%%%%%%%%%%%
%%%%%%%%%%%%%%%%%%%%%%%%%%%%%%%%%%%%%%%%%%%%%%%%%%%%%%%%%%%%%%%%%%%%%%%%%%%%%%%%%%%%%%%%%%%%%%%%%%%%%

The first moment of the kinetic equation (\ref{RTA}) reads
\begin{eqnarray}
\int\!\! dP \; p^\nu  p\cdot\partial f =  \frac{1}{\tau_{\rm eq}} \int\!\! dP \,p^\nu \,  p\cdot U
 \left(  f_{\rm eq} -f \right).
\label{fm1}
\end{eqnarray}
With the energy-momentum tensors defined by the second moments of the distribution functions,
\begin{eqnarray}
T^{\mu\nu} = \int dP p^\mu p^\nu f,  \quad 
T^{\mu\nu}_{\rm eq} = \int dP p^\mu p^\nu f_{\rm eq},
\label{Tmunus}
\end{eqnarray} 
we may rewrite Eq.~(\ref{fm1}) as
\begin{eqnarray}
\partial_\mu T^{\mu\nu} = \frac{1}{\tau_{\rm eq}} \left( U_\mu T^{\mu\nu}_{\rm eq} - U_\mu T^{\mu\nu} \right).
\label{fm2}
\end{eqnarray}
Since we want to conserve energy and momentum in the system, the left-hand side of Eq.~(\ref{fm2}) must vanish
\begin{eqnarray}
\partial_\mu T^{\mu\nu} = 0.
\label{enmomcon}
\end{eqnarray}
This leads us to the conclusion that the first-moment equations (\ref{fm1}) and (\ref{fm2}) are satisfied only if the Landau matching condition is satisfied
\begin{eqnarray}
U_\mu T^{\mu\nu}_{\rm eq} =  U_\mu T^{\mu\nu}.
\label{LM1}
\end{eqnarray}

The form of the distribution function (\ref{fxi}) implies that the energy-momentum tensor of the anisotropic system has the structure
\begin{eqnarray}
T^{\mu \nu} = \varepsilon\, U^\mu U^\nu + P_X X^\mu X^\nu + P_Y Y^\mu Y^\nu + P_Z Z^\mu Z^\nu
= \varepsilon\, U^\mu U^\nu + \sum_I P_I I^\mu I^\nu,
\label{Tmunu}
\end{eqnarray}
where $\varepsilon$ is the energy density, while $P_X, P_Y$ and $P_Z$ are three different pressure components. In the local rest frame the energy-momentum tensor has the diagonal structure,
\begin{equation}
T^{\mu \nu} =  \left(
\begin{array}{cccc}
\varepsilon & 0 & 0 & 0 \\
0 & P_X & 0 & 0 \\
0 & 0 & P_Y & 0 \\
0 & 0 & 0 & P_Z
\end{array} \right).
\label{Tmunuarray}
\end{equation}
In local equilibrium, $\varepsilon = \varepsilon_{\rm eq}$ and the three pressures become equal, $P_X = P_Y = P_Z = P_{\rm eq} = \varepsilon/3 $. Hence, the equilibrium energy-momentum tensor has the expected form
\begin{eqnarray}
T^{\mu \nu}_{\rm eq} &=& 
\varepsilon_{\rm eq} U^\mu U^\nu + P_{\rm eq} X^\mu X^\nu + P_{\rm eq} Y^\mu Y^\nu + P_{\rm eq} Z^\mu Z^\nu, \nonumber \\
&=& \varepsilon_{\rm eq} U^\mu U^\nu - P_{\rm eq}
\Delta^{\mu\nu} = \left(\varepsilon_{\rm eq} +
P_{\rm eq} \right) U^\mu U^\nu - P_{\rm eq} g^{\mu\nu}.
\label{Tmunueq}
\end{eqnarray}
The use of the expressions (\ref{Tmunu}) and (\ref{Tmunueq}) in the Landau matching condition (\ref{LM1}) leads directly to the two equations
\begin{eqnarray}\label{p_e_d}
\varepsilon U^\mu = \varepsilon_{\rm eq} U^\mu, \quad  \quad  \varepsilon = \varepsilon_{\rm eq}.
\end{eqnarray}
We thus see that the Landau matching condition implies simply that the energy density of the system should be equal to the energy density of the thermal background. This requirement allows us to determine the effective temperature $T$ appearing in the thermal distribution $f_{\rm eq}$.

For boost-invariant and cylindrically symmetric systems only two out of four equations appearing in the conservation laws (\ref{enmomcon}) are independent. They are the same as those derived in Ref. \cite{Florkowski:2011jg} and may be written in the compact form as
\begin{eqnarray}
 D \varepsilon + \varepsilon \, \theta  - \sum_I P_I \theta_I = 0 
 \label{enmom1}
\end{eqnarray}
and
\begin{eqnarray}
&& \left( X\cdot\partial \right) P_X  + P_X \left( \partial\cdot X \right) - \varepsilon \left( X\cdot DU \right)  - P_Y \left[ X\cdot\left( Y\cdot\partial \right)Y \right] - P_Z \left[ X\cdot\left( Z\cdot\partial \right)Z \right] = 0. \label{enmom2}
\end{eqnarray}
See Sec.~\ref{sect:explicitr} for the explicit formulas of the derivatives appearing in (\ref{enmom1}) and (\ref{enmom2}).

The standard dissipative hydrodynamics is based on the gradient expansion around the isotropic background. In this case, one  usually considers small deviations from the equilibrium values. From this point of view it is interesting and useful to consider the close-to-equilibrium limit of our framework. Therefore, we introduce deviations from the equilibrium pressure, $\pi_I$'s, defined by the relations
\begin{equation}\label{pi_I}
P_X = P_{\rm eq} + \pi_X, 
\qquad P_Y = P_{\rm eq} + \pi_Y,
\qquad P_Z = P_{\rm eq} + \pi_Z.
\end{equation}
The sum of the pressure deviations is equal to zero
\begin{eqnarray}
\sum_I \pi_I = \pi_X + \pi_Y + \pi_Z = 0.
\label{sumpiI}
\end{eqnarray}
The equilibrium pressure $P_{\rm eq}$ is one third of the energy density, $P_{\rm eq}=\varepsilon/3$. Changing from $P_I$'s to $\pi_I$'s we rewrite Eq.~(\ref{enmom1}) in the equivalent forms
\begin{equation}\label{alenmom2}
 D\varepsilon + \frac{4}{3}\theta - \sum_I \pi_I\theta_I = 0, \qquad  D\ln\varepsilon = -\frac{4}{3}\theta + \sum_I \frac{\pi_I}{\varepsilon}\,\theta_I.
\end{equation}

%%%%%%%%%%%%%%%%%%%%%%%%%%%%%%%%%%%%%%%%%%%%%%%%%%%%%%%%%%%%%%%%%%%%%%%%%%%%%%%%%%%%%%%%%%%%%%%%%%%%%
%%%%%%%%%%%%%%%%%%%%%%%%%%%%%%%%%%%%%%%%%%%%%%%%%%%%%%%%%%%%%%%%%%%%%%%%%%%%%%%%%%%%%%%%%%%%%%%%%%%%%
\subsection{Close-to-equilibrium behavior}
\label{sect:closetoeq}
%%%%%%%%%%%%%%%%%%%%%%%%%%%%%%%%%%%%%%%%%%%%%%%%%%%%%%%%%%%%%%%%%%%%%%%%%%%%%%%%%%%%%%%%%%%%%%%%%%%%%
%%%%%%%%%%%%%%%%%%%%%%%%%%%%%%%%%%%%%%%%%%%%%%%%%%%%%%%%%%%%%%%%%%%%%%%%%%%%%%%%%%%%%%%%%%%%%%%%%%%%%

In order to find the pressure deviations $\pi_I$'s in the close-to-equilibrium limit, we expand the anisotropic distribution function around the thermal background,
\begin{equation}
f \simeq f_{\rm eq} \left( 1 + \frac{\lambda - T}{ T^2 } (p\cdot U) - \frac{ \xi_X (p\cdot X)^2  + \xi_Y (p\cdot Y)^2 + \xi_Z (p\cdot Z)^2 }{2T (p\cdot U)} \right).
\end{equation}
Here, we neglect higher order contributions in $\xi_I$'s and in the difference $\lambda-T$. Then, the energy-momentum tensor reads
\begin{eqnarray}
 T^{\mu\nu} &\simeq& T^{\mu\nu}_{\rm eq} + 96\, \pi\,k\,T^3\,U^\mu U^\nu \left( \frac{}{}  \lambda - T \right)- 32\,\pi\,k\,T^3\,\Delta^{\mu\nu}\left( \frac{}{} \lambda - T \right)
 \nonumber \\
 &&  -\frac{32\, \pi}{5}k\, T^4 \left(  \frac{}{} \xi_X X^\mu X^\nu + \xi_Y Y^\mu Y^\nu + \xi_Z Z^\mu Z^\nu \right).
\end{eqnarray}
Using the Landau matching~(\ref{p_e_d}) we find that $\lambda=T$ in the leading order. Hence, the energy-momentum tensor in the leading order reads
\begin{equation}
 T^{\mu\nu} \simeq T^{\mu\nu}_{\rm eq} -\frac{32 \pi}{5}k\, T^4 \left(  \frac{}{} \xi_X X^\mu X^\nu + \xi_Y Y^\mu Y^\nu + \xi_Z Z^\mu Z^\nu \right).
\end{equation}
This expression helps us to identify directly the pressure corrections:
\begin{equation}\label{shear-eq1}
 \pi_X \simeq - \frac{ 32 \pi k T^4}{5}\xi_X,
 \qquad  \pi_Y \simeq - \frac{32 \pi k T^4}{5}\xi_Y,
  \qquad 
 \pi_Z \simeq - \frac{32 \pi k T^4}{5}\xi_Z.
\end{equation}
It is interesting to observe that the pressure corrections are directly proportional to the anisotropy parameters. 

%%%%%%%%%%%%%%%%%%%%%%%%%%%%%%%%%%%%%%%%%%%%%%%%%%%%%%%%%%%%%%%%%%%%%%%%%%%%%%%%%%%%%%%%%%%%%%%%%%%%%
%%%%%%%%%%%%%%%%%%%%%%%%%%%%%%%%%%%%%%%%%%%%%%%%%%%%%%%%%%%%%%%%%%%%%%%%%%%%%%%%%%%%%%%%%%%%%%%%%%%%%
\section{Energy density and anisotropic pressure}
\label{sect:enedenaniso}
%%%%%%%%%%%%%%%%%%%%%%%%%%%%%%%%%%%%%%%%%%%%%%%%%%%%%%%%%%%%%%%%%%%%%%%%%%%%%%%%%%%%%%%%%%%%%%%%%%%%%
%%%%%%%%%%%%%%%%%%%%%%%%%%%%%%%%%%%%%%%%%%%%%%%%%%%%%%%%%%%%%%%%%%%%%%%%%%%%%%%%%%%%%%%%%%%%%%%%%%%%%

The energy density for the anisotropic distribution (\ref{fxi}) may be obtained from the contraction of the energy-momentum tensor with the four-vectors $U$, namely, $\varepsilon = U_\mu U_\nu T^{\mu\nu}$. This gives
\begin{eqnarray}
\varepsilon(\lambda,\xi) = 24 \pi k \lambda^4 {\cal R}(\xi),
\label{eps1}
\end{eqnarray}
where the function ${\cal R}(\xi)$ is defined by the integral (for details see Sec.~\ref{sect:R})
\begin{equation}\label{R}
 {\cal R}(\xi) = \frac{1}{4\pi \sqrt{ \prod_J(1 + \xi_J) } }\int_0^{2\pi}\!\!\!\! {\rm d}\phi \int_0^\pi \!\!\! {\rm d}\theta \sin\theta \sqrt{ \frac{ \cos^2\phi\sin^2\theta }{1+\xi_X} + \frac{ \sin^2\phi\sin^2\theta }{1+\xi_Y} + \frac{ \cos^2\theta }{1+\xi_Z} }.
\end{equation}
In the case of local equilibrium we have
\begin{eqnarray}
\varepsilon_{\rm eq}(T) = 24 \pi k T^4.
\label{epseq1}
\end{eqnarray}
This leads us to the equivalent formulation of the Landau matching condition (\ref{LM1}) in the form
\begin{eqnarray}
T^4 = \lambda^4 {\cal R}(\xi),
\label{LM2}
\end{eqnarray}
which resembles the condition derived first in \cite{Martinez:2010sc}. It is important to notice, however, that our definitions of the parameters $\lambda$ and $\xi$ are different from the definitions of the parameters $\Lambda$ and $\xi$ used in \cite{Martinez:2010sc}~\footnote{The use of a single parameter $\xi$ in \cite{Martinez:2010sc} corresponds to the use of the non-zero parameter $\zeta_Z$ in our approach with the constraint $\zeta_X=\zeta_Y=0$. Our triplet of $\xi_I$'s satisfies the condition (\ref{sumofxis}). By the way, one can use the condition (\ref{sumofxis}) to check that the expansion of the function ${\cal R}(\xi)-1$ around zero has only quadratic terms in $\xi_I$'s. This is in agreement with the leading order result $\lambda=T$.}.

Similarly to the energy density, the three components of anisotropic pressure may be obtained from three contractions of the energy-momentum tensor with the four-vectors $X$, $Y$, and $Z$, namely, $P_X = X_\mu X_\nu T^{\mu\nu}$, $P_Y = Y_\mu Y_\nu T^{\mu\nu}$, and $P_Z = Z_\mu Z_\nu T^{\mu\nu}$. This leads to the formula
\begin{eqnarray}
P_I(\lambda,\xi) = 24 \pi k \lambda^4 {\cal H}_I (\xi),
\label{P_I}
\end{eqnarray}
where the functions ${\cal H}_I$ may be obtained by differentiation of the function ${\cal R}$ (the definitions of the functions ${\cal H}_I$ as integrals, which leads directly to (\ref{H_I}), are given also in Sec.~\ref{sect:R})
\begin{eqnarray}\nonumber
 {\cal H}_I &=& -\frac{ 2\left( 1+ \xi_I  \right) }{\sqrt{\prod_J (1 + \xi_J)}} \partial_{\xi_I}\left[ \sqrt{\prod_J (1 + \xi_J)}\;{\cal R} \right] \\
 &=& -2\left( 1+ \xi_I  \right){\cal R} \,  \partial_{\xi_I}\left\{  \ln \left[ \sqrt{\prod_J (1 + \xi_J)} \; {\cal R} \right] \right\} 
 \label{H_I} \\ \nonumber
 &=& -{\cal R} -2{\cal R}\left( 1 + \xi_I \right) \partial_{\xi_I} \left[ \ln\left( \frac{}{} {\cal R} \right) \right]	.
\end{eqnarray}
Since we consider a system of massless particles, $\varepsilon = P_X +P_Y + P_Z$ and the functions ${\cal H}_I$ satisfy the constraint
\begin{equation}
\sum_I {\cal H}_I = {\cal R}.
\label{sumH_I}
\end{equation}

%%%%%%%%%%%%%%%%%%%%%%%%%%%%%%%%%%%%%%%%%%%%%%%%%%%%%%%%%%%%%%%%%%%%%%%%%%%%%%%%%%%%%%%%%%%%%%%%%%%%%
%%%%%%%%%%%%%%%%%%%%%%%%%%%%%%%%%%%%%%%%%%%%%%%%%%%%%%%%%%%%%%%%%%%%%%%%%%%%%%%%%%%%%%%%%%%%%%%%%%%%%
\section{Second moment of kinetic equation}
\label{sect:2mom}
%%%%%%%%%%%%%%%%%%%%%%%%%%%%%%%%%%%%%%%%%%%%%%%%%%%%%%%%%%%%%%%%%%%%%%%%%%%%%%%%%%%%%%%%%%%%%%%%%%%%%
%%%%%%%%%%%%%%%%%%%%%%%%%%%%%%%%%%%%%%%%%%%%%%%%%%%%%%%%%%%%%%%%%%%%%%%%%%%%%%%%%%%%%%%%%%%%%%%%%%%%%

The second moment of the Boltzmann equation may be written in the form analogous to Eq.~(\ref{fm2}),
\begin{eqnarray}
\partial_\lambda \Theta^{\lambda\mu\nu} = \frac{1}{\tau_{\rm eq}} \left(U_\lambda\Theta_{\rm eq}^{\lambda\mu\nu} - U_\lambda\Theta^{\lambda\mu\nu}\right),
 \label{tmom}
\end{eqnarray}
where 
\begin{eqnarray}
\Theta^{\lambda\mu\nu} = \int\!\! dP \; p^\lambda p^\mu p^\nu f, \quad
\Theta^{\lambda\mu\nu}_{\rm eq} = \int\!\! dP \; p^\lambda p^\mu p^\nu f_{\rm eq}.
\label{Thetas}
\end{eqnarray}
The only non-vanishing terms in (\ref{Thetas}) are those with an even number of each spatial index. In the covariant form they read
\begin{eqnarray} 
\Theta &=& \Theta_U \left[ U\otimes U \otimes U\right] 
\nonumber \\
&& \,+\, \Theta_X \left[ U\otimes X \otimes X +X\otimes U \otimes X + X\otimes X \otimes U\right] 
\nonumber \\ 
&& \,+\,  \Theta_Y  \left[ U\otimes Y \otimes Y +Y\otimes U \otimes Y + Y\otimes Y \otimes U\right]
\nonumber \\
&& \,+\, \Theta_Z \left[ U\otimes Z \otimes Z +Z\otimes U \otimes Z + Z\otimes Z \otimes U\right].
 \label{Theta}
\end{eqnarray}
Due to the mass-shell condition $p^2 = m^2 = 0$, the coefficients in the expansion (\ref{Theta}) are not independent. One may check that
\begin{eqnarray}
\Theta_X + \Theta_Y + \Theta_Z = \Theta_U.
\label{ThetaU}
\end{eqnarray}
This and other tensor identities may be most easily checked in the local rest frame. A similar argument holds for the projections of $\Theta_{\rm eq}^{\lambda \mu\nu}$. In addition, due to the rotation invariance of the equilibrium distribution we have
\begin{eqnarray}
\Theta^{\rm eq}_X = \Theta^{\rm eq}_Y = \Theta^{\rm eq}_Z = \Theta_{\rm eq}\,.
\label{ThetaU}
\end{eqnarray}
Out of the ten independent equations in (\ref{tmom}) five are trivial $0=0$ equations. They correspond to the contractions of (\ref{tmom}) with  $U\otimes Y$, $U\otimes Z$, $X\otimes Y$, $X\otimes Z$, and $Y\otimes Z$. The contraction with $U\otimes U$ may be represented as a linear combination of the contractions with $X\otimes X$,
$Y\otimes Y$, and $Z\otimes Z$. As a consequence, we deal with four independent contractions, namely, with $U\otimes X$, $X\otimes X$, $Y\otimes Y$, and $Z\otimes Z$. The contraction of (\ref{tmom}) with $U\otimes X$ gives
\begin{eqnarray}
D \left(\Theta_U + 2 \Theta_X \right) +
\left( X\cdot\partial \right) \Theta_X = 
 \frac{\sinh\theta_\perp}{\tau} \left( \Theta_Z - \Theta_X \right) +
 \frac{\cosh\theta_\perp}{r} \left( \Theta_Y - \Theta_X \right),
\label{tmomUX}
\end{eqnarray}
with $X\otimes X$ 
\begin{eqnarray}
D \Theta_X + \Theta_X  \left( \theta - 2\theta_X \right) = \frac{1}{\tau_{\rm eq}} \left(\Theta_{\rm eq} - \Theta_X \right),
\label{tmomXX}
\end{eqnarray}
with $Y\otimes Y$ 
\begin{eqnarray}
 D \Theta_Y + \Theta_Y  \left( \theta - 2\theta_Y \right) = \frac{1}{\tau_{\rm eq}} \left(\Theta_{\rm eq} - \Theta_Y \right),
\label{tmomYY}
\end{eqnarray}
and, finally, with $Z\otimes Z$
\begin{eqnarray}
 D \Theta_Z + \Theta_Z  \left( \theta - 2\theta_Z \right) = \frac{1}{\tau_{\rm eq}} \left(\Theta_{\rm eq} - \Theta_Z \right).
\label{tmomZZ}
\end{eqnarray}
%

%%%%%%%%%%%%%%%%%%%%%%%%%%%%%%%%%%%%%%%%%%%%%%%%%%%%%%%%%%%%%%%%%%%%%%%%%%%%%%%%%%%%%%%%%%%%%%%%%%%%
\section{Selection of equations of motion - matching with Israel-Stewart theory}
\label{sect:set}
%%%%%%%%%%%%%%%%%%%%%%%%%%%%%%%%%%%%%%%%%%%%%%%%%%%%%%%%%%%%%%%%%%%%%%%%%%%%%%%%%%%%%%%%%%%%%%%%%%%%%

In the considered model we have five independent parameters (more precisely, five scalar functions of the proper time, $\tau$, and the transverse distance, $r$). These are: the momentum scale, $\lambda$, the effective temperature, $T$, the transverse rapidity, $\theta_\perp$, and two independent anisotropy parameters, for example, $\xi_X$ and $\xi_Y$. The two nontrivial equations from the first moment of the Boltzmann equation, Eqs.~(\ref{enmom1}) and (\ref{enmom2}), as well as  the Landau matching condition~(\ref{LM2}) ensure local energy and momentum conservation. Therefore, these three equations should be definitely included in the computational scheme of anisotropic hydrodynamics.

The problem arises which equations should be taken into account in addition to the first-moment equations. We need two extra equations and they should be selected out of Eqs.~(\ref{zm3}), (\ref{tmomUX}), (\ref{tmomXX}), (\ref{tmomYY}), and (\ref{tmomZZ}). An important requirement for our approach is that it must agree with the Israel-Stewart approach in the close-to-equilibrium limit. In this case, the pressure corrections satisfy the three symmetric equations~\footnote{Due to the conditions $\sum_I \pi_I =0$ and $\sum_I \sigma_I =0$ only two out of three equations in (\ref{2_ord_visc}) are independent.}
\begin{equation}
 \tau_\pi D\pi_I + \pi_I = 2 \eta \sigma_I + F_\eta \,\pi_I,
 \label{2_ord_visc}
\end{equation}
where \cite{Muronga:2003ta}
\begin{eqnarray}
F_\eta &=& -  \eta T \partial \cdot \left( \frac{\alpha_1}{2 T} U \right) . %  -  \eta T \partial \cdot \left( \frac{\tau_\pi}{2\eta T} U \right) .
\label{F}
\end{eqnarray}
In Eqs.~(\ref{2_ord_visc}) and (\ref{F}) the quantity $\tau_\pi$ is the relaxation time for the shear viscous corrections $\pi_I$, $\eta$ is the shear viscosity, and $\alpha_1$ is one of the kinetic coefficients appearing in second order hydrodynamics~\cite{Israel:1979wp}.  The symmetric form of the three equations appearing in (\ref{2_ord_visc}) suggests that one should use Eqs.~(\ref{tmomXX}), (\ref{tmomYY}), and (\ref{tmomZZ}) as a starting point for possible generalizations of (\ref{2_ord_visc}) to the case of high pressure anisotropy. The use of the zeroth moment equation combined with one of the equations obtained from the second moment leads to asymmetric treatment of different anisotropies, which contradicts the symmetric form of Eqs.~(\ref{2_ord_visc}).

In the remaining part of this Section we show that two linear combinations of  Eqs.~(\ref{tmomXX})--(\ref{tmomZZ}) provide indeed a system of equations which agree with Eqs.~(\ref{2_ord_visc}) in the close-to-equilibrium limit. At first, it is useful to take advantage of the fact that $\Theta_I$'s are  positive,
\begin{equation}
  \Theta_I = \int\!\! dP \,  p \, (p\cdot I)^2 f =  \frac{ 32 \, \pi \, k \, \lambda^5 }{ \sqrt{ \prod_J(1+\xi_J) } } \frac{1}{1+\xi_I},
\label{Theta_I}
\end{equation}
\begin{equation}
  \Theta_{\rm eq} = \int\!\! dP \,  p \, (p\cdot I)^2 f_{\rm eq} =  { 32 \, \pi \, k \, T^5 } \qquad  (I=X,Y,Z).
\label{Theta_eq}  
\end{equation}
Then, we rewrite Eqs.~(\ref{tmomXX})--(\ref{tmomZZ}) dividing each of them first by $\Theta_I$. In this way we obtain
\begin{eqnarray}
 D \ln\Theta_I + \theta - 2\theta_I = \frac{1}{\tau_{\rm eq}} \left[ \frac{\Theta_{\rm eq}}{\Theta_I} - 1 \right].
\label{(I)/I}
\end{eqnarray}
In the next step, we define the two desired equations by taking Eqs.~(\ref{(I)/I}) for $I=X$ and $I=Y$, and by subtracting one third of the sum of Eqs.~(\ref{(I)/I}) from these two equations
\begin{eqnarray}
 D \ln\Theta_I + \theta - 2\theta_I -\frac{1}{3}\sum_J \left[\frac{}{} D \ln\Theta_J + \theta - 2\theta_J\right]  = \frac{1}{\tau_{\rm eq}} \left[ \frac{\Theta_{\rm eq}}{\Theta_I} - 1 \right] -\frac{1}{3}\sum_J \left\{ \frac{1}{\tau_{\rm eq}} \left[ \frac{\Theta_{\rm eq}}{\Theta_J} - 1 \right]  \right\} \quad (I=X,Y).
\label{sum}
\end{eqnarray}
Of course, other choices of the two indices $I$ are also possible. The very important feature of our strategy is that fulfilling Eqs.~(\ref{sum}) for arbitrary two indices implies that the same equation is fulfilled for the remaining third index. To demonstrate this property, we first make use of Eqs.~(\ref{Theta_I}) and (\ref{Theta_eq}) to rewrite Eqs.~(\ref{sum}) in the simpler form as
\begin{equation}
 \frac{D\xi_I}{1+\xi_I} 
 -\frac{1}{3}\sum_J\frac{D\xi_J}{1+\xi_J}
 + 2\sigma_I + \frac{\xi_I}{\tau_{\rm eq}} \left( \frac{T}{\lambda} \right)^5\sqrt{ \prod_J(1+\xi_J) }  =  0 \quad (I=X,Y).
\label{explicit-sum2}
\end{equation}
If Eq.~(\ref{explicit-sum2}) is fulfilled for $I=X$ and $I=Y$, the same equation holds for $I=Z$. Indeed, if we use the properties $\sigma_Z = -\sigma_X - \sigma_Y$ and $\xi_Z = -\xi_X - \xi_Y$, then the straightforward calculation shows
\begin{eqnarray}\nonumber
 && \frac{D\xi_Z}{1+\xi_Z} 
 -\frac{1}{3}\sum_J\frac{D\xi_J}{1+\xi_J} 
 + 2\sigma_Z + \frac{\xi_Z}{\tau_{\rm eq}} \left( \frac{T}{\lambda} \right)^5\sqrt{ \prod_J(1+\xi_J) }  \\\nonumber
 && = \frac{D\xi_Z}{1+\xi_Z} 
-\frac{1}{3}\sum_J\frac{D\xi_J}{1+\xi_J}  
 + \left[ -2 \sigma_X -\frac{\xi_X}{\tau_{\rm eq}} \left( \frac{T}{\lambda} \right)^5\sqrt{ \prod_J(1+\xi_J) }  \right] + \left[ -2 \sigma_Y -\frac{\xi_Y}{\tau_{\rm eq}} \left( \frac{T}{\lambda} \right)^5\sqrt{ \prod_J(1+\xi_J) }  \right]  \\
 && = \frac{D\xi_Z}{1+\xi_Z} + \frac{D\xi_X}{1+\xi_Z} + \frac{D\xi_Y}{1+\xi_Y}  -  \sum_J\frac{D\xi_J}{1+\xi_J} \equiv 0.
\end{eqnarray}

Close to equilibrium, the anisotropy parameters $\xi_I$ are proportional to the pressure corrections $\pi_I$. Using Eqs.~(\ref{p_e_d}), (\ref{shear-eq1}), and (\ref{epseq1}), we find
\begin{equation}\label{shear-eq2}
  \xi_X \simeq -\frac{15}{4} \frac{ \pi_X}{ \varepsilon }  \qquad \xi_Y \simeq -\frac{15}{4} \frac{ \pi_Y}{ \varepsilon } \qquad \xi_Z \simeq -\frac{15}{4} \frac{ \pi_Z}{ \varepsilon }.
\end{equation}
Since $\xi_X + \xi_Y + \xi_Z =0$ and $\lambda=T$ up to quadratic terms in the anisotropy parameters, Eqs.~(\ref{explicit-sum2}) reads
\begin{equation}\label{IS-equivalent}
 D\xi_I + 2\sigma_I + \frac{\xi_I}{\tau_{\rm eq}}  \simeq  0.
\end{equation}
Using Eqs.~(\ref{shear-eq2}) and multiplying the last equation by $-4\varepsilon/15$ we obtain
\begin{equation}\label{IS-equivalent}
 D\pi_I -\pi_I D\ln\varepsilon - 2 \left( \frac{4}{15}\, \varepsilon \right)\sigma_I + \frac{\pi_I}{\tau_{\rm eq}}  \simeq  0,  
\end{equation}
or
\begin{equation}\label{IS-equivalent2}
\tau_{\rm eq} D\pi_I + \pi_I = 2  \left( \frac{4}{15}\, \varepsilon \, \tau_{\rm eq}\right) \sigma_I +\tau_{\rm eq}\, \pi_I \,  D\ln\varepsilon.
\end{equation}
The two terms on the left-hand side of (\ref{IS-equivalent2}) and the first term on the right-hand side of (\ref{IS-equivalent2}) agree with the corresponding terms in Eq.~(\ref{2_ord_visc}) provided we connect the relaxation time $\tau_\pi$ and the shear viscosity $\eta$ with the relaxation time $\tau_{\rm eq}$ and the energy density $\varepsilon$ by the expressions
\begin{equation}\label{transport_coefficients}
 \tau_\pi = \tau_{\rm eq}, \qquad \eta = \frac{4}{15} \, \varepsilon \, \tau_{\rm eq}.
\end{equation}
Finally, one may find the value of the coefficient $\alpha_1$ comparing the last term with the expression for  $ F_\eta \pi_I$%check that the last term in Eq.~(\ref{2_ord_visc}) is equivalent to $ F_\eta \pi_I$
. Using (\ref{F}), one finds

\begin{eqnarray}
 \pi_I F_\eta &=& %- \pi_I  \eta T \partial \cdot \left( \frac{\tau_\pi}{2\eta T} U \right)  = -\pi_I \left[ \frac{1}{2}D\tau_\pi  + \frac{1}{2}\tau_\pi \, \theta -\frac{1}{2}\tau_\pi D\ln\left( \eta \, T \right) \right].
 - \pi_I  \eta T \, \partial \cdot \left( \frac{\alpha_1}{2 T} U \right) = \pi_I \left[ -\partial\cdot\left(\frac{}{}\eta\alpha_1\,  U\right) + \eta\alpha_1D\ln\left(\frac{}{}\eta T\right) \right] = \\ \nonumber
 &=& \pi_I \left[ -\left( \eta\alpha_1 \right)\theta  - D\left( \eta\alpha_1 \right) +\left(\eta\alpha_1\right)D\ln\left(\frac{}{}\eta T\right) \right]
\end{eqnarray}
With the relaxation time $\tau_\pi$ and the viscosity $\eta$ identified through Eq.~(\ref{transport_coefficients}) and also using Eqs.~(\ref{p_e_d}), (\ref{alenmom2}), and (\ref{epseq1}) we obtain
\begin{eqnarray}\nonumber
 \pi_I F_\eta &=& %-\pi_I \left[ \frac{1}{2}D\tau_{\rm eq}  -\frac{3}{8}\tau_{\rm eq}D\ln\varepsilon +\frac{3}{8}\tau_{\rm eq}\sum_J \frac{\pi_J}{\varepsilon}\theta_J  -\frac{1}{2}\tau_\pi D\ln\varepsilon -\frac{1}{2}D\tau_{\rm eq}-\frac{1}{8}\tau_{\rm eq} D\ln\varepsilon \right]  \\
 %&=& \pi_I \, \tau_{\rm eq} \, D\ln \varepsilon - \frac{3}{8}\tau_{\rm eq}\left[  \sum_J \frac{\pi_J}{\varepsilon}\theta_J\right]\pi_I.
 \pi_I \left[ \frac{3}{4}\left(\eta\alpha_1\right)D\ln\varepsilon -\frac{1}{4}\left(\eta\alpha_1\right) \sum_J \frac{\pi_J}{\varepsilon}\theta_J - D\left(\eta\alpha_1\right) +\frac{5}{4}\left(\eta\alpha_1\right)D\ln\varepsilon + \left(\eta\alpha_1\right) D\ln\tau_\pi \right] = \\
 &=& \pi_I \left[ 2\left(\eta\alpha_1\right) D\ln\varepsilon +\left(\eta\alpha_1\right)D\ln\tau_\pi -D\left(\eta\alpha_1\right) \right]
\end{eqnarray}
Since the last term is quadratic in the pressure corrections, it can be neglected, while the remaining part equals the last term in Eq.~(\ref{IS-equivalent2}) when%. 

\begin{equation}
\eta\alpha_1  = \frac{1}{2} \tau_\pi \qquad \hbox{or} \qquad \alpha_1=\frac{\tau_\pi}{2\eta},
\end{equation}
which is consistent with the Israel-Stewart theory~\cite{Muronga:2001zk}.

%%%%%%%%%%%%%%%%%%%%%%%%%%%%%%%%%%%%%%%%%%%%%%%%%%%%%%%%%%%%%%%%%%%%%%%%%%%%%%%%%%%%%%%%%%%%%%%%%%%%
\section{Entropy source}
\label{sect:ent}
%%%%%%%%%%%%%%%%%%%%%%%%%%%%%%%%%%%%%%%%%%%%%%%%%%%%%%%%%%%%%%%%%%%%%%%%%%%%%%%%%%%%%%%%%%%%%%%%%%%%%

An important verification of our scheme based on 
Eqs.~(\ref{enmom1}), (\ref{enmom2}), (\ref{LM2}), and (\ref{explicit-sum2}) is checking if it leads to the positively defined entropy source. Using the Boltzmann definition of the entropy current, we find that it is proportional to the particle density
\begin{eqnarray}
\sigma^\mu = \sigma U^\mu = 4 n U^\mu,
\label{entcur}
\end{eqnarray}
where $n$ is defined by Eq.~(\ref{n}), hence
\begin{eqnarray}
\sigma(\lambda,\xi) = 
\frac{32\pi k \lambda^3}{\sqrt{1+\xi_X} \sqrt{1+\xi_Y}\sqrt{1+\xi_Z}}.
\label{sigma}
\end{eqnarray}
Combining this equation with the formula for the energy density we find
\begin{equation}\label{eps2}
 \varepsilon = 24\, \pi \, k \left( \frac{ \sigma }{32\, \pi \, k} \sqrt{ \prod_I \left( 1 +\xi_I \right) } \right)^{\frac{4}{3}} {\cal R} = \frac{3}{8} \left( 4\, \pi \, k \right)^{-\frac{1}{3}} \sigma^{\frac{4}{3}} \left[ \prod_J \left( 1 + \xi_J \right) \right]^{\frac{2}{3}} {\cal R}.
\end{equation}
Substituting this equation into Eq.~(\ref{enmom1}) gives
\begin{equation}
 \frac{4}{3} \left( \frac{}{} D\ln\sigma  + \theta \right) +\frac{2}{3}\sum_I \frac{ D \xi_I }{ 1 + \xi_I } + D\ln {\cal R} -\sum_I \frac{ \pi_I }{\varepsilon} \theta_I = 0,
 \label{entr_sour_1}
\end{equation}
We note that the expression in the bracket on the left-hand side of Eq.~(\ref{entr_sour_1}) is proportional to the entropy source
\begin{eqnarray}
\Sigma = \partial_\mu \sigma^\mu =
\partial_\mu \left(\sigma U^\mu \right) = D \sigma + \sigma \theta.
\end{eqnarray}

We shall express now the last two terms in (\ref{entr_sour_1}) in terms of the functions ${\cal R}$ and ${\cal H}_I$. From Eq.~(\ref{H_I}) we calculate the $\xi_I$ derivative of $\ln({\cal R})$ 
\begin{equation}\label{derR}
\partial_{\xi_I} \left[ \ln\left( \frac{}{} {\cal R} \right) \right]  = -\frac{1}{2\left( 1 +\xi_I \right)}\left[ 1 + \frac{ {\cal H}_I }{ {\cal R} } \right].
\end{equation}
Hence, the convective derivative $D\ln({\cal R})$ reads
\begin{equation}\label{DlnR}
 D\ln\left( \frac{}{} {\cal R} \right) = \sum_I D\xi_I \, \partial_{\xi_I} \left[ \ln\left( \frac{}{} {\cal R} \right) \right] =  -\frac{1}{2} \sum_I \left[ 1 +\frac{ {\cal H}_I }{ {\cal R} } \right] \frac{ D\xi_I }{ 1 +\xi_I }.
\end{equation}
On the other hand, using definitions of the pressure corrections~(\ref{pi_I}) and of the functions ${\cal H}_I$, we find a useful expression for the $\pi/\varepsilon$ ratio
\begin{equation}
 \frac{\pi_I}{\varepsilon} = -\frac{1}{3}\left[ 1 -3 \frac{ {\cal H}_I  }{ {\cal R} } \right] 
 \label{pi_I/eps}
\end{equation}
and
\begin{equation}
 \sum_I \frac{\pi_I}{\varepsilon} \theta_I = -\frac{1}{3}\sum_I \left[ 1 -3 \frac{ {\cal H}_I  }{ {\cal R} } \right] \left( \frac{1}{3}\theta + \theta_I -\frac{1}{3}\theta \right) = -\frac{1}{6} \sum_I \left[ 1 -3 \frac{ {\cal H}_I  }{ {\cal R} } \right] 2\sigma_I.
 \label{sumpi_I/eps}
\end{equation}
Here we replaced the components of the expansion tensor $\theta_I$ by the components of the shear tensor $\sigma_I$.

Using Eqs.~(\ref{derR})--(\ref{sumpi_I/eps}), which are exact and do not refer to the small anisotropy limit, we may write
\begin{equation}\label{entr_sour_2}
 \frac{4}{3} \, \frac{ \partial_\mu \sigma^\mu }{ \sigma } +\frac{1}{6}\sum_I \left[ 1 - 3\frac{ {\cal H}_I }{ {\cal R} } \right] \frac{ D \xi_I }{ 1 + \xi_I }  +\frac{1}{6}\sum_I \left[ 1 -  3\frac{ {\cal H}_I }{ {\cal R} } \right] 2 \sigma_I = 0,
\end{equation}
or, equivalently,

\begin{equation}\label{entr_sour_3}
 \frac{ \partial_\mu \sigma^\mu }{ \sigma } = - \sum_I\left[ \frac{1}{8} - \frac{3}{8} \frac{ {\cal H}_I }{ {\cal R} } \right] \left( \frac{ D\xi_I }{ 1 + \xi_I }  + 2 \sigma_I \right).
\end{equation}
Using now Eqs.~(\ref{LM2}), (\ref{sumH_I}), and (\ref{explicit-sum2}) we find
\begin{equation}\label{entr_sour_4}
 \frac{ \partial_\mu \sigma^\mu }{ \sigma } = \frac{1}{\tau_{\rm eq}}{\cal R}^{\frac{5}{4}}\sqrt{\prod_J\left( 1 + \xi_J \right)} \, \sum_I\left[ \frac{1}{8} - \frac{3}{8} \frac{ {\cal H}_I }{ {\cal R} } \right] \xi_I =  -\frac{3}{ 8 \tau_{\rm eq} }{\cal R}^{\frac{5}{4}}\sqrt{\prod_J\left( 1 + \xi_J \right)} \, \sum_I\frac{ {\cal H}_I }{ {\cal R} }\xi_I \ge 0.
\end{equation}
The last inequality has been checked numerically in the allowed range of the parameters $\xi_X$ and $\xi_Y$, see Eqs.~(\ref{sumofxis}) and (\ref{range}). For small anisotropies we use Eqs.~(\ref{shear-eq2}) and (\ref{pi_I/eps}) to find that
\begin{eqnarray}
 \frac{ \partial_\mu \sigma^\mu }{ \sigma } = \frac{1}{10 \tau_{\rm eq}} \sum_I \xi_I^2,
\label{entsorceapp} 
\end{eqnarray}
which is again consistent with the Israel-Stewart theory.

%%%%%%%%%%%%%%%%%%%%%%%%%%%%%%%%%%%%%%%%%%%%%%%%%%%%%%%%%%%%%%%%%%%%%%%%%%%%%%%%%%%%%%%%%%%%%%%%%%%%
\section{Summary and conclusions}
\label{sect:con}
%%%%%%%%%%%%%%%%%%%%%%%%%%%%%%%%%%%%%%%%%%%%%%%%%%%%%%%%%%%%%%%%%%%%%%%%%%%%%%%%%%%%%%%%%%%%%%%%%%%%%

In this paper we have used the projection method for boost-invariant and cylindrically symmetric systems to introduce a new formulation of anisotropic hydrodynamics that allows for three different values of pressure acting in three different  directions. Our considerations have been based on the Boltzmann kinetic equation with the collision term treated in the relaxation time approximation. The momentum anisotropy has been included explicitly in the leading term of the distribution function. 

A novel feature of our work is the complete analysis of the second moment of the Boltzmann equation, in addition to the zeroth and first moments that have been analyzed in earlier studies. The framework of anisotropic hydrodynamics should include five equations for five unknown functions: $\lambda$, $T$, $\theta_\perp$, $\xi_X$ and $\xi_Y$. The first two equations follow from the energy and momentum conservation, Eqs.~(\ref{enmom1}) and (\ref{enmom2}). Their explicit, extended versions are
\begin{eqnarray}
&& \left( \cosh \theta_\perp \partial_\tau 
+ \sinh \theta_\perp \partial_r \right) 
\varepsilon(\lambda,\xi) 
+ \varepsilon(\lambda,\xi) \left[ \cosh \theta_\perp \left( \frac{1}{\tau} + \partial_r \theta_\perp \right) + \sinh \theta_\perp \left( 
\frac{1}{r} + \partial_\tau \theta_\perp \right) \right] \nonumber \\
&& + P_X(\lambda,\xi)  \left( \cosh \theta_\perp \partial_r \theta_\perp + \sinh \theta_\perp \partial_\tau \theta_\perp \right) + P_Y(\lambda,\xi)  \frac{\sinh \theta_\perp}{r} +P_Z(\lambda,\xi)  \frac{\cosh \theta_\perp}{\tau} = 0,
 \label{fineq1}
\end{eqnarray}
and
\begin{eqnarray}
&& \left( \sinh \theta_\perp \partial_\tau 
+ \cosh \theta_\perp \partial_r \right) P_X(\lambda,\xi)     + \varepsilon \left( \sinh \theta_\perp \partial_r \theta_\perp 
+ \cosh \theta_\perp \partial_\tau \theta_\perp  \right)  \nonumber \\
&& + P_X(\lambda,\xi)  \left[ \sinh \theta_\perp \left( \frac{1}{\tau} + \partial_r \theta_\perp \right) + \cosh \theta_\perp \left( 
\frac{1}{r} + \partial_\tau \theta_\perp \right) \right] - P_Y(\lambda,\xi)  \frac{\cosh \theta_\perp}{r} - P_Z(\lambda,\xi)  \frac{\sinh \theta_\perp}{\tau} = 0. \label{fineq2}
\end{eqnarray}
The main result of the present work is that Eqs.~(\ref{fineq1}) and (\ref{fineq2}) should be supplemented with the two equations obtained from the second moment of the Boltzmann equation
\begin{eqnarray}
&& \frac{1}{1+\xi_I} \left( \cosh \theta_\perp \partial_\tau 
+ \sinh \theta_\perp \partial_r \right) \xi_I
 -\frac{1}{3}\sum_J\frac{1}{1+\xi_J}
 \left( \cosh \theta_\perp \partial_\tau 
+ \sinh \theta_\perp \partial_r \right) \xi_J
 \nonumber \\
&& + 2\sigma_I + \frac{\xi_I}{\tau_{\rm eq}} \left( \frac{T}{\lambda} \right)^5\sqrt{ \prod_J(1+\xi_J) }  =  0 \qquad (I=X,Y).
\label{fineq34}
\end{eqnarray}
The effective temperature appearing in (\ref{fineq34}) should be obtained from the Landau matching condition which, for the sake of convenience, we also repeat here
\begin{eqnarray}
\left(\frac{T}{\lambda}\right)^4 =  {\cal R}(\xi).
\label{fineq5}
\end{eqnarray}
The numerical analysis of Eqs.~(\ref{fineq1})--(\ref{fineq5}) is left for a separate study.

\bigskip

{\bf Acknowledgments}: L.T. and W.F. were supported in part by the Polish National Science Center grants with decisions No. DEC-2012/06/A/ST2/00390 and No.
DEC-2012/05/B/ST2/02528, respectively.

%%%%%%%%%%%%%%%%%%%%%%%%%%%%%%%%%%%%%%%%%%%%%%%%%%%%%%%%%%%%%%%%%%%%%%%%%%%%%%%%%%%%%%%%%%%%%%%%%%%%%
%%%%%%%%%%%%%%%%%%%%%%%%%%%%%%%%%%%%%%%%%%%%%%%%%%%%%%%%%%%%%%%%%%%%%%%%%%%%%%%%%%%%%%%%%%%%%%%%%%%%%
\section{Appendix: Explicit formulas for derivatives}
\label{sect:explicitr}
%%%%%%%%%%%%%%%%%%%%%%%%%%%%%%%%%%%%%%%%%%%%%%%%%%%%%%%%%%%%%%%%%%%%%%%%%%%%%%%%%%%%%%%%%%%%%%%%%%%%%
%%%%%%%%%%%%%%%%%%%%%%%%%%%%%%%%%%%%%%%%%%%%%%%%%%%%%%%%%%%%%%%%%%%%%%%%%%%%%%%%%%%%%%%%%%%%%%%%%%%%%

The total time (or convective) derivative, $D = U^\alpha \partial_\alpha = U \cdot \partial$, describes the change of a physical quantity in the local rest frame. In the remaining part of this Section we collect the formulas involving $D$ and other derivatives which are useful in dealing with  hydrodynamic equations.

\medskip \noindent
Directional derivatives:
\begin{eqnarray}
U \cdot \partial &=&
\cosh\theta_\perp \partial_\tau 
+ \sinh\theta_\perp \partial_r, 
\quad Y \cdot \partial = \frac{1}{r} \partial_\phi,
\nonumber \\
X \cdot \partial &=& \sinh\theta_\perp \partial_\tau 
+ \cosh\theta_\perp \partial_r, 
\quad
Z \cdot \partial = \frac{1}{\tau} \partial_{\eta_\parallel}.
\label{useful-eqns-1}
\end{eqnarray}

\medskip \noindent
Divergencies:
\begin{eqnarray}
\partial \cdot U &=&
\cosh\theta_\perp \left(\frac{1}{\tau} + \partial_r \theta_\perp \right)
+ \sinh\theta_\perp \left( \frac{1}{r} + \partial_\tau \theta_\perp \right),  \quad
\partial \cdot Y = 0,
\nonumber \\
\partial \cdot X &=&
\sinh\theta_\perp \left(\frac{1}{\tau} + \partial_r \theta_\perp \right)
+ \cosh\theta_\perp \left( \frac{1}{r} + \partial_\tau \theta_\perp \right),  \quad
\partial \cdot Z = 0.
\label{useful-eqns-2}
\end{eqnarray}

\medskip \noindent
Convective derivatives of $U$, $X$, $Y$, and $Z$:
\begin{eqnarray}
D U  = (U \cdot \partial) U &=&
X \left(\cosh\theta_\perp \partial_\tau \theta_\perp + \sinh\theta_\perp \partial_r \theta_\perp \right),  \quad
D Y = (U \cdot \partial) Y = 0,
\nonumber \\
D X = (U \cdot \partial) X &=&
U \left(\cosh\theta_\perp \partial_\tau \theta_\perp + \sinh\theta_\perp \partial_r \theta_\perp \right),  \quad
D Z = (U \cdot \partial) Z = 0.
\label{useful-eqns-3}
\end{eqnarray}

\medskip \noindent
Directional derivatives of $U$, $X$, $Y$, and $Z$:
\begin{eqnarray}
(X \cdot \partial) U &=&
X \left(\sinh\theta_\perp \partial_\tau \theta_\perp + \cosh\theta_\perp \partial_r \theta_\perp \right),  \quad
(X \cdot \partial) Y = 0,
\nonumber \\
(X \cdot \partial) X &=&
U \left(\sinh\theta_\perp \partial_\tau \theta_\perp + \cosh\theta_\perp \partial_r \theta_\perp \right),  \quad
(X \cdot \partial) Z = 0,
\label{useful-eqns-4}
\end{eqnarray}
\begin{eqnarray}
(Y \cdot \partial) U &=&
\frac{\sinh\theta_\perp}{r} \, Y,  \quad
(Y \cdot \partial) Y = \frac{1}{r}\left( \frac{}{} \sinh\theta_\perp U - \cosh\theta_\perp X \right),
\nonumber \\
(Y \cdot \partial) X &=&
\frac{\cosh \theta_\perp}{r} \, Y,  \quad
(Y \cdot \partial) Z =0,
\label{useful-eqns-5}
\end{eqnarray}
\begin{eqnarray}
(Z \cdot \partial) U &=&
\frac{\cosh\theta_\perp}{\tau} \, Z,  \quad
(Z \cdot \partial) Y = 0,
\nonumber \\
(Z \cdot \partial) X &=&
\frac{ \sinh\theta_\perp}{\tau} \, Z ,  \quad
(Z \cdot \partial) Z = \frac{1}{\tau}\left( \frac{}{} \cosh\theta_\perp U - \sinh\theta_\perp X \right).
\label{useful-eqns-6}
\end{eqnarray}

%%%%%%%%%%%%%%%%%%%%%%%%%%%%%%%%%%%%%%%%%%%%%%%%%%%%%%%%%%%%%%%%%%%%%%%%%%%%%%%%%%%%%%%%%%%%%%%%%%%%%
%%%%%%%%%%%%%%%%%%%%%%%%%%%%%%%%%%%%%%%%%%%%%%%%%%%%%%%%%%%%%%%%%%%%%%%%%%%%%%%%%%%%%%%%%%%%%%%%%%%%%
\section{Appendix: Integrals for energy density and pressure}
\label{sect:R}
%%%%%%%%%%%%%%%%%%%%%%%%%%%%%%%%%%%%%%%%%%%%%%%%%%%%%%%%%%%%%%%%%%%%%%%%%%%%%%%%%%%%%%%%%%%%%%%%%%%%%
%%%%%%%%%%%%%%%%%%%%%%%%%%%%%%%%%%%%%%%%%%%%%%%%%%%%%%%%%%%%%%%%%%%%%%%%%%%%%%%%%%%%%%%%%%%%%%%%%%%%%

In order to pass from Eq.~(\ref{entr_sour_1}) to Eq.~(\ref{entr_sour_2}) we need several properties of the function ${\cal R}$ defined in~(\ref{eps1}). They follow most easily from the representation of ${\cal R}$ in the local rest frame, 
\begin{eqnarray}\nonumber
 \varepsilon &=& \int dP (p\cdot U)^2 \, f = k\int \frac{ {\rm d}^3 {\bf p} }{p} p^2 \exp\left[-\frac{1}{\lambda} \sqrt{ \sum_I (p^I)^2 \left( \frac{}{} 1 +\xi_I \right) } \right]  \\
 &=& \frac{ 6\, k\, \lambda^4 }{ \sqrt{ \prod_I (1 +\xi_I) } } \int_0^{2\pi}\!\!\!\! {\rm d}\phi \int_0^\pi \!\!\! {\rm d}\theta \sin\theta \sqrt{ \frac{ \cos^2\phi\sin^2\theta }{1+\xi_X} + \frac{ \sin^2\phi\sin^2\theta }{1+\xi_Y} + \frac{ \cos^2\theta }{1+\xi_Z} } = 24 \pi k \lambda^4 {\cal R}.
\end{eqnarray}
For simplicity of notation we use here the symbol $p^I$ to denote the three-momentum component $p_i$. 

\bigskip

Similar expressions may be found for the pressures $P_X$, $P_Y$ and $P_Z$ 
\begin{eqnarray}
 P_X &=& \int dP \, (p\cdot X)^2 \, f = k\int \frac{ {\rm d}^3 {\bf p} }{p} \left( p^X \right)^2 \exp\left[-\frac{1}{\lambda} \sqrt{ \sum_I (p^I)^2 \left( \frac{}{} 1 +\xi_I \right) } \right]  \\ \label{H_X}
 &=& \frac{ 6\, k\, \lambda^4 }{ \sqrt{ \prod_I (1 +\xi_I) } } \int_0^{2\pi}\!\!\!\! {\rm d}\phi \int_0^\pi \!\!\! {\rm d}\theta \sin\theta \frac{ \cos^2\phi \sin ^2\theta }{\left( 1 + \xi_X \right) \sqrt{ \frac{ \cos^2\phi\sin^2\theta }{1+\xi_X} + \frac{ \sin^2\phi\sin^2\theta }{1+\xi_Y} + \frac{ \cos^2\theta }{1+\xi_Z} } } = 24\, \pi \, k \, \lambda^4\; {\cal H}_X.  \nonumber
\end{eqnarray}
 
\begin{eqnarray}
P_Y &=& \int dP \, (p\cdot Y)^2 \, f = k\int \frac{ {\rm d}^3 {\bf p} }{p} \left( p^Y \right)^2 \exp\left[-\frac{1}{\lambda} \sqrt{ \sum_I (p^I)^2 \left( \frac{}{} 1 +\xi_I \right) } \right]  \\ \label{H_Y}
 &=& \frac{ 6\, k\, \lambda^4 }{ \sqrt{ \prod_I (1 +\xi_I) } } \int_0^{2\pi}\!\!\!\! {\rm d}\phi \int_0^\pi \!\!\! {\rm d}\theta \sin\theta \frac{ \sin^2\phi \sin ^2\theta }{\left( 1 + \xi_Y \right) \sqrt{ \frac{ \cos^2\phi\sin^2\theta }{1+\xi_X} + \frac{ \sin^2\phi\sin^2\theta }{1+\xi_Y} + \frac{ \cos^2\theta }{1+\xi_Z} } } = 24\, \pi \, k \, \lambda^4\; {\cal H}_Y .
\end{eqnarray} 
 
\begin{eqnarray} 
P_Z &=& \int dP \, (p\cdot Z)^2 \, f = k\int \frac{ {\rm d}^3 {\bf p} }{p} \left( p^Z \right)^2 \exp\left[-\frac{1}{\lambda} \sqrt{ \sum_I (p^I)^2 \left( \frac{}{} 1 +\xi_I \right) } \right] = \\ 
\label{H_Z}
 &=& \frac{ 6\, k\, \lambda^4 }{ \sqrt{ \prod_I (1 +\xi_I) } } \int_0^{2\pi}\!\!\!\! {\rm d}\phi \int_0^\pi \!\!\! {\rm d}\theta \sin\theta \frac{ \cos^2\theta }{\left( 1 + \xi_Z \right) \sqrt{ \frac{ \cos^2\phi\sin^2\theta }{1+\xi_X} + \frac{ \sin^2\phi\sin^2\theta }{1+\xi_Y} + \frac{ \cos^2\theta }{1+\xi_Z} } } = 24\, \pi \, k \, \lambda^4\; {\cal H}_Z.
\end{eqnarray}
The equations above define the functions ${\cal H}_I$.

\end{document}